\documentclass[final, 5p, twocolumn, 10pt, a4paper, times, sort&compress]{elsarticle}

\pdfoutput=1

\makeatletter
\def\ps@pprintTitle{%
  \let\@oddhead\@empty
  \let\@evenhead\@empty
  \def\@oddfoot{\reset@font\hfil\thepage\hfil}
  \let\@evenfoot\@oddfoot
}
\makeatother

\usepackage{amsmath} 
\usepackage{amssymb}
\usepackage{algorithmic}
\usepackage{fixltx2e}
\usepackage{graphicx}
\usepackage[caption=false,font=small]{subfig}
\usepackage{setspace}
\usepackage[retainorgcmds]{IEEEtrantools}
\usepackage[final, colorlinks]{hyperref}

\hypersetup{pdftitle={Reconstruction-classification method for quantitative photoacoustic tomography}}

\hypersetup{pdfauthor={Emma~R.~Malone, Samuel~Powell, Ben~T.~Cox, Simon~R.~Arridge}}

\hypersetup{pdfsubject={Quantitative photoacoustic tomography}}

\hypersetup{pdfkeywords={quantitative photoacoustic tomography, image reconstruction, chromophores}}

\newcommand{\mua}{{\mu_a}}
\newcommand{\mus}{{\mu_s}}
\newcommand{\musp}{{\mu'_s}}
\newcommand{\p}{\mathrm{p}}
\newcommand{\rd}{\mathrm{d}}

\begin{document} 
\begin{frontmatter}
\title{Reconstruction-classification method for quantitative photoacoustic tomography} 

\author[mpb]{\corref{cor1}Emma~R.~Malone}
\ead{e.malone@ucl.ac.uk}
\author[mpb,cs]{Samuel~Powell}
\author[mpb]{Ben~T.~Cox}
\author[cs]{Simon~R.~Arridge}
\cortext[cor1]{Corresponding author}

\address[mpb]{Department of Medical Physics and Biomedical Engineering, University College London, Gower Street, LONDON, WC1E 6BT, UK.}
\address[cs]{Department of Computer Science, University College London, Gower Street, LONDON, WC1E 6BT, UK.}

\begin{abstract}
We propose a combined reconstruction-classification method for simultaneously recovering absorption and scattering in turbid media from images of absorbed optical energy. This method exploits knowledge that optical parameters are determined by a limited number of classes to iteratively improve their estimate. Numerical experiments show that the proposed approach allows for accurate recovery of absorption and scattering in 2 and 3 dimensions, and delivers superior image quality with respect to traditional reconstruction-only approaches.
\end{abstract}


\end{frontmatter}

\section{Introduction}
\label{sect:intro} 

Photoacoustic tomography (PAT) is an emerging technique for in vivo imaging of soft biological tissue \cite{Beard2011}. This hybrid modality uses ultrasound to detect optical contrast, combining the high resolution of acoustic methods with the spectroscopic capability of optical imaging. To generate a PA image, a short laser pulse is shone into the object, the ultrasonic waves emitted following the heating of the tissue are measured, and an image of the absorbed optical energy field is recovered. Whereas purely optical methods suffer from poor spatial resolution, acoustic waves propagate with minimal scattering and PAT can achieve 100 micron resolution at depths of several centimetres. However, PA images provide only qualitative information about the tissue, and are not directly related to tissue morphology and functionality. The principal difficulty is that the PA image is the product of both the optical absorption coefficient (which is directly related to underlying tissue composition) and the light distribution (which is not). This severely restricts the range of applications for which PAT is suitable.  

Quantitative photoacoustic tomography (QPAT) aims to provide clinically valuable images of the optical absorption and scattering coefficients, or chromophore (light-absorbing molecules) concentrations from conventional PA images via an image reconstruction method \cite{Cox2012}.  A model of light propagation is required to relate the absorbed optical energy to the light fluence and tissue parameters. The primary challenge of QPAT is solving the non-linear imaging problem. In particular, recovering the scattering coefficient is especially difficult to due to its weak dependence on the absorbed energy density.

In this paper, we develop a method for solving the image reconstruction problem for QPAT by alternating reconstruction and segmentation steps in an automated iterative process. We introduce a probabilistic model that describes optical properties in terms of a limited number of optically distinct classes, which may correspond to tissues or chromophores. These are identified and characterized by a classification, or segmentation, algorithm. This approach allows for the use of information retrieved by the classification in the reconstruction stage, and vice versa. The aim of the reconstruction is to choose solutions for which the image parameters take values close to a finite set of discrete points. The aim of the classification algorithm is to progressively improve the parametric optical model, and correct for errors in the initial assumptions. Multinomial models have been employed previously in the related fields Diffuse Optical Tomography \cite{Hiltunen2009} and Electrical Impedance Tomography \cite{Malone2015}. For QPAT, the main advantage is that this approach enables accurate recovery of both the absorption and scattering coefficients, simultaneously.
\section{Numerical methods}
\subsection{Quantitative photoacoustic imaging}

A conventional PAT image is proportional to the absorbed optical energy
\begin{equation}
H(\boldsymbol{r}) = \hat{\Gamma}(\boldsymbol{r}) \mua(\boldsymbol{r}) \phi\left(\mua\left(\boldsymbol{r}\right), \musp\left(\boldsymbol{r}\right)\right)\ \ \ \ \ \ \ \ \boldsymbol{r}\in \Omega,
\end{equation}
where  $\boldsymbol{r}$ is a position vector within the domain $\Omega$, $\mua$ and $\musp$ are the optical absorption and reduced scattering coefficients, $\phi$ is the optical fluence, and $\hat{\Gamma}$ is the Gr\"uneisen parameter. The Gr\"uneisen parameter represents the efficiency with which the tissue converts heat into acoustic pressure, and is often taken to be constant $\hat{\Gamma}(\boldsymbol{r})=1, \forall\boldsymbol{r}\in\Omega$. The fluence is dependent on the optical parameters and illumination pattern in the whole domain. The problem of recovering the optical parameters $\left(\mua,\musp\right)$ from a conventional PAT image is known as the \emph{quantitative} problem. The optical absorption $\mua$ is of particular interest because it is fundamentally related to underlying tissue physiology and functionality, and encodes clinically useful information such as tissue oxygenation levels and chromophore concentrations. Conversely, the absorbed energy density $H$ depends non-trivially on optical absorption, thus is not directly related to tissue morphology because it is distorted, structurally and spectrally, by the non-uniform light fluence.

\subsection{The diffusion model of light transport}
In order to recover the optical parameters $\left(\mua,\musp\right)$, a model of light propagation within the tissue is required. For highly scattering media and far from boundaries and sources, a low order spherical harmonic approximation to the \emph{radiative transfer equation} is suitable. The \emph{diffusion approximation} is given by \cite{Arridge1999}
\begin{equation}
\left(\mua - \nabla\cdot \kappa(\boldsymbol{r})\nabla\right)\phi(\boldsymbol{r})=q(\boldsymbol{r}),
\label{eq_diffapproxcont}
\end{equation}
where $q(\boldsymbol{r})$ is an isotropic source term, and $\kappa= 1/3\musp$ is the diffusion coefficient.

We set Robin boundary conditions
\begin{equation}
\phi(\boldsymbol{r})+\frac{1}{2A}\kappa(\boldsymbol{r})\hat{n}\cdot\nabla\phi(\boldsymbol{r})=0 \ \ \ \ \ \boldsymbol{r}\in\delta\Omega
\end{equation}
where $A$ accounts for the refractive index mismatch at the boundary. 

\subsection{Minimization-based QPAT imaging}
In this paper, we adopt a gradient-based minimization approach to image reconstruction. Typically, both $\mua$ and $\musp$ are unknown and need to be recovered simultaneously from the absorbed energy density. An objective function is defined, which measures the distance between the conventional PAT image $H^m$ and the data predicted by the model for the current estimates $H(\mua,\musp)$.
\begin{equation}
\mathcal{E} = \frac{1}{2}\int_\Omega{(H^m-H(\mua,\musp))^2\mathrm{d}\Omega}.
\label{eq_objfuncont}
\end{equation}
In order to treat the problem for a generic geometry, the Finite Element Method (FEM) is employed, whereby a weak formulation of the diffusion approximation \eqref{eq_diffapproxcont} is considered. A discretization of the domain is defined, and the fluence and optical parameters are expressed in terms of piecewise linear basis functions $u_i(\boldsymbol{r})$: $\chi\approx\sum_i\chi_iu_i(\boldsymbol{r})$ for $\chi\in\left\{\mua,\musp,\phi\right\}$, where $\chi_i$ are nodal coefficients and $i=1,\dots,N$. 

We assume that the data $\boldsymbol{d}^m$ is the absorbed energy density $H^m$, projected onto a particular basis $\left\{\Psi_j\right\}$,
\begin{align}
\boldsymbol{d}^m &= \left\{d^m_j, j = 1,\ldots, N\right\},\\
d^m_j &= \int_\Omega{H^m(\boldsymbol{r})\Psi_j(\boldsymbol{r})\mathrm{d}\Omega} = \left<\Psi,\ H^m\right>.
\end{align}
Choices for $\left\{\Psi_j\right\}$ include:
\begin{enumerate}
\item Point sampling $\Psi_j(\boldsymbol{r}) = \delta (\boldsymbol{r}-\boldsymbol{r}_j)$,
\item Piecewise-linear sampling $\Psi_j = u_j$,
\item Sinc sampling $\Psi_j = \mathrm{sinc}(\left|\boldsymbol{r}-\boldsymbol{r}_j\right|)$.
\end{enumerate}
Substituting into the  the objective function \eqref{eq_objfuncont} leads to the discrete form of the objective function
\begin{equation}
\mathcal{E} = \frac{1}{2}\sum_i(d_j^m - \left<{\Psi_i,\ H(\mua,\musp)}\right>)^2 =  \frac{1}{2}\sum_i(d_j^m - \left<\Psi_j,\ \mua\phi\right>)^2.
\label{eq_objfunnoreg}
\end{equation}

If a single illumination source is used and both absorption and scattering are undetermined, the problem is ill posed \cite{Cox2012}. In this study, the non-uniqueness of the solution was removed by using multiple illumination patterns \cite{Bal2011}, thus the objective function must be summed over the number of sources. In the following, we have omitted this sum for ease of notation. Prior information regarding the solution can be included by adding a regularization term 
\begin{equation}
\mathcal{E} = \frac{1}{2}\sum_j(d_j^m - \left<\Psi_j,\ \mua\phi\right>)^2 + \mathcal{R}(\boldsymbol{\mua},\boldsymbol{\musp}).
\label{eq_objfungen}
\end{equation}
In the Bayesian framework, an image is obtained by maximizing the posterior probability of the parameters, given the data:
\begin{equation}
\p(\boldsymbol{\mua},\boldsymbol{\musp}|\boldsymbol{d}^m)\propto\p(\boldsymbol{d}^m|\boldsymbol{\mua},\boldsymbol{\musp})\p(\boldsymbol{\mua},\boldsymbol{\musp}).
\end{equation}
Under this interpretation, the regularization term $\mathcal{R}$ is given by the negative log of the prior probability distribution
\begin{equation}
\mathcal{R}(\boldsymbol{\mua},\boldsymbol{\musp}) = -\log \p(\boldsymbol{\mua},\boldsymbol{\musp}).
\label{eq_reg}
\end{equation}

\subsection{Gradient calculations}
Cox \textit{et al.} \cite{Cox2007} have shown that, for the continuous case, the gradient of \eqref{eq_objfuncont} with respect to $\mua$ at position $\boldsymbol{r}^0$ is given by 
\begin{equation}
\left.\frac{\partial \mathcal{E}}{\partial \mua}\right|_{\boldsymbol{r}^0} = -\left.\phi\left(H^m-H\right)\right|_{\boldsymbol{r}^0}+\left.\phi\cdot\phi^*\right|_{\boldsymbol{r}^0},
\end{equation}
where $\phi^*$ is the \emph{adjoint} light field. In the following, we derive the expression for the gradient in the discrete case. 

The sampled forward model can be expressed as a vector $\boldsymbol{H} = \left\{H_j, j = 1,\ldots, N\right\}$,
\begin{IEEEeqnarray}{rCl}
H_j & = & \int_\Omega{H(\boldsymbol{r})\Psi_j(\boldsymbol{r})\mathrm{d}\Omega} = \left<\Psi,\ H^m\right>,\nonumber\\
& = & \sum_{ik}\mua_{\,i}\phi_k\int_{\Omega}{\Psi_j(\boldsymbol{r})u_i(\boldsymbol{r})u_k(\boldsymbol{r})\mathrm{d}\Omega} = \boldsymbol{\phi}^T \mathbf{C}^j\boldsymbol{\mua},
\label{eq_Hdiscr}
\end{IEEEeqnarray}
where $\mathbf{C}^j$ is a sparse matrix with entries $i,\,k$ where the support of the basis functions $\Psi_j(\boldsymbol{r}),\,u_i(\boldsymbol{r}),\,u_k(\boldsymbol{r})$ overlap. Taking the derivative of \eqref{eq_objfunnoreg} with respect to $\mua_{\,i}$, we have
\begin{equation}
\frac{\partial \mathcal{E}}{\partial \mua_{\,i}}= -\sum_j\left(\frac{\partial H_j}{\partial \mua_{\,i}}\right)(d_j^m-H_j).
\label{eq_grad1}
\end{equation}
Using the expression for the absorbed energy density \eqref{eq_Hdiscr},
\begin{equation}
\frac{\partial H_j}{\partial \mua_{\,i}}=  \boldsymbol{e}_i^T\mathbf{C}^j\boldsymbol{\phi} + \boldsymbol{\mua}^T\mathbf{C}^j\frac{\partial\boldsymbol{\phi}}{\partial \mua_{\,i}},
\end{equation}
where $\boldsymbol{e}_i$ is a vector of zeros with a single 1 in position $i$. Substituting into \eqref{eq_grad1} gives
\begin{equation}
\frac{\partial \mathcal{E}}{\partial \mua_{\,i}}= - \sum_j ( \boldsymbol{e}_i^T\mathbf{C}^j\boldsymbol{\phi} + \boldsymbol{\mua}^T\mathbf{C}^j\frac{\partial\boldsymbol{\phi}}{\partial \mua_{\,i}})(d^m_j-H_j).
\label{eq_dedmua1}
\end{equation}
The first term in equation \eqref{eq_dedmua1} is 
\begin{IEEEeqnarray}{rCl}
 \sum_j {\boldsymbol{e}_i^T\mathbf{C}^j\boldsymbol{\phi}(d^m_j-H_j)} & = &  \sum_{j,i,k} e_i C^j_{ik}\phi_k(d^m_j-H_j)\nonumber\\
&=&  \sum_{j,k} \phi_k (d^m_j-H_j) \int_{\Omega}{\Psi_j(\boldsymbol{r})u_i(\boldsymbol{r})u_k(\boldsymbol{r})\mathrm{d}\Omega} \nonumber\\
&=&  \boldsymbol{\phi}^T \mathbf{E}^i (\boldsymbol{d}^m-\boldsymbol{H})
\end{IEEEeqnarray}
where $\mathbf{E}^i$ is given by a reordering of $C^j_{ik}$
\begin{equation}
E^i_{kj} = \int_{\Omega}{\Psi_j(\boldsymbol{r})u_i(\boldsymbol{r})u_k(\boldsymbol{r})\mathrm{d}\Omega}.
\end{equation}
Note that while $\mathbf{C}^j$ is symmetric, in general $\mathbf{E}^i$ is not.

It remains to determine $\frac{\partial\boldsymbol{\phi}}{\partial \mua_{\,i}}$. The discrete form of the DA model \eqref{eq_diffapproxcont} assumes the form \cite{Schweiger2014c}:
\begin{equation}
\left(\mathbf{M}+\mathbf{K}+\mathbf{F}\right)\boldsymbol{\phi} =\boldsymbol{Q},
\label{eq_diffapproxdiscr}
\end{equation}
where 
\begin{equation}
M_{jk}=\sum_i\mua_{\,i}\int_{\Omega}{u_iu_ju_k\rd \Omega},
\end{equation}
\begin{equation}
K_{jk}=\sum_i \kappa_i \int_{\Omega}{u_i \nabla u_j \cdot \nabla u_k \rd \Omega},
\end{equation}
\begin{equation}
F_{jk}=\sum_i \frac{1}{2A} \int_{\partial\Omega}{u_j u_k \rd S},
\end{equation}
\begin{equation}
Q_{j}=\sum_i q_i \int_{\Omega}{u_i u_j \rd \Omega}.
\end{equation}
Taking the derivative of equation \eqref{eq_diffapproxdiscr} with respect to the $i$th coefficient of $\boldsymbol{\mua}$,
\begin{equation}
\left(\mathbf{M}+\mathbf{K}+\mathbf{F}\right)\frac{\partial\boldsymbol{\phi}}{\partial\mua_{\,i}} = - \mathbf{V}^i_{\mua}\boldsymbol{\phi}
\label{eq_dphi}
\end{equation}
where
\begin{equation}
V_{\mua,\,jk}^i = \int_{\Omega}{u_iu_ju_k\rd \Omega}
\end{equation}
is given by the derivative of the system matrix. We define the adjoint field $\boldsymbol{\phi}^*$ as the solution to the equation 
\begin{equation}
\left(\mathbf{M}+\mathbf{K}+\mathbf{F}\right)\boldsymbol{\phi}^* =\boldsymbol{Q}^*
\label{eq_adj}
\end{equation}
where
\begin{equation}
\boldsymbol{Q}^* = \sum_j\boldsymbol{\mua}^T\mathbf{C}^j(d^m_j-H_j)
\end{equation}
is the adjoint source. Taking $\boldsymbol{\phi}^*\cdot$ \eqref{eq_dphi} $-\ \frac{\partial\boldsymbol{\phi}}{\partial \mua_{\,i}}\ \cdot$ \eqref{eq_adj} we obtain
\begin{equation}
\sum_j \boldsymbol{\mua}^T\mathbf{C}^j\frac{\partial\boldsymbol{\phi}}{\partial \mua_{\,i}}(d^m_j-H_j)  =  -{\boldsymbol{\phi}}^T  \mathbf{V}^i_{\mua}\boldsymbol{\phi}^*.
\end{equation}
Substituting into \eqref{eq_dedmua1} gives the expression for the derivative with respect to $\mua_{\,i}$
\begin{equation}
\frac{\partial \mathcal{E}}{\partial \mua_{\,i}} =  \boldsymbol{\phi}^T (\mathbf{V}^i_{\mua}\boldsymbol{\phi}^* - \mathbf{E}^i (\boldsymbol{d}^m-\boldsymbol{H})).
\end{equation}
The derivative with respect to $\musp_{\,i}$ can be derived analogously:
\begin{equation}
\frac{\partial \mathcal{E}}{\partial \musp_{\,i}} =  - \frac{\partial\kappa_i}{\partial\musp_{\,i}}{\boldsymbol{\phi}}^T  \mathbf{V}^i_{\musp}\boldsymbol{\phi}^*,
\end{equation}
where
\begin{equation}
V^i_{\musp,\,jk} = \int_{\Omega}{u_i \nabla u_j \cdot \nabla u_k\rd \Omega},
\end{equation}
and $\frac{\partial\kappa_i}{\partial\musp_{\,i}} = -1/3\musp_{\,i}^2$. Note that calculation of the gradient only requires two runs of the forward model. The forward problem was solved using the Toast++ software package \cite{Schweiger2014c}.

Choosing point-sampling $\Psi_j(\boldsymbol{r}) = \delta (\boldsymbol{r}-\boldsymbol{r}_j)$, gives simply $\mathbf{C}^j\,=\,\mathbf{E}^i\,=\,\mathbf{I}$. In this study, we chose piecewise-linear sampling $\Psi_j = u_j$, so we had $\mathbf{C}^j\,=\,\mathbf{E}^i\,=\,\mathbf{V}^i_{\mua}$ and 
\begin{equation}
\frac{\partial \mathcal{E}}{\partial \mua_{\,i}} =  \boldsymbol{\phi}^T \mathbf{V}^i_{\mua} (\boldsymbol{\phi}^* - \boldsymbol{d}^m +\boldsymbol{H}).
\end{equation}

\section{Reconstruction-classification method for QPAT}
A reconstruction-classification scheme is devised, which enables the recovery $\mua$ and $\musp$ by approaching the image reconstruction and segmentation problems simultaneously. At each reconstruction step, we minimize a regularized objective function, where the regularization term is given by a mixture model. At each classification step, the result of the previous reconstruction step is employed to update the class parameters for the multinomial model. We alternate between reconstruction and classification steps for a fixed number of iterations.

\subsection{Mixture model for $\mua$ and $\musp$}
In this section we introduce a probability model for $\mua$ and $\musp$, which encodes prior knowledge about the optical parameters and allows us to bias the solution of the imaging problem accordingly. We assume that an array of labels $\boldsymbol{\zeta}_i$ can be determined for each node, such that 
\begin{equation}
\zeta_{ij}=\left\{
\begin{array}{ll}
1& \text{if the } i \text{th node is assigned to the } j \text{th class};\\
0& \text{otherwise.}\\
\end{array}\right.
\end{equation}
The labels constitute \emph{hidden variables} on which the image parameters are dependant. For each class $j=1,\ldots,J$, a mean vector $\boldsymbol{m}_j=\left(\bar{\mua}_{\,j},\bar{\mus}'_{\,j}\right) \in \mathbb{R}^{2}$ is defined, and the closeness of the optical parameters to the mean values is described by a covariance matrix $\Sigma_j \in \mathbb{R}^{2\times2}$. 

We assume that if $\zeta_{ij}=1$, the probability distribution for $\boldsymbol{x}_i=\left(\mua_{\,i},\musp_{\,i}\right)$ is given by a multivariate Gaussian distribution
\begin{equation}
\p(\boldsymbol{x}_i|\boldsymbol{\theta}_j) = \mathcal{N}(\boldsymbol{m}_j, \Sigma_j),
\end{equation}
where $\boldsymbol{\theta}_j$ indicates the set of class parameters $(\boldsymbol{m}_j, \Sigma_j)$. 

The prior probability distribution of the class properties $\boldsymbol{\theta}_j$ is given by the conjugate prior to the Gaussian distribution. Prior information about the distribution of the class means or covariances can be encoded by choosing the parameters of the conjugate prior accordingly. Using a non-informative prior for the class means we have $\p(\boldsymbol{m}_{j})\propto 1$. The conjugate prior distribution for the covariance of a normal distribution is given by the \emph{normal inverse Wishart distribution}
\begin{equation}
\mathrm{NIW}(\nu_j,\Gamma_j)=\left|\Sigma_j\right|^{-(\nu+d+1)/2}\exp\left[-\frac{1}{2}\mathrm{Tr}(\Gamma_j\Sigma_j^{-1})\right],
\end{equation}
where $d$ is the dimension of the domain, $\nu_j$ indicates the number of degrees of freedom, and $\Gamma_j$ is a scaling matrix. If the prior is non-informative, then $\nu_j=0$ and $\Gamma_j=0$, and the probability distribution of the class parameters becomes
\begin{equation}
\p(\boldsymbol{\theta}_j)\propto\left|\Sigma_j\right|^{-(d+1)/2},
\end{equation}
which is known as Jeffreys prior.

The probability that the set of labels $\boldsymbol{\zeta}_{i}=\{\zeta_{i1},...,\zeta_{ij},...,\zeta_{iJ}\}$ is assigned to the $i$th node is given by a multinomial distribution
\begin{equation}
\p(\boldsymbol{\zeta}_{i}|\boldsymbol{\lambda})=\prod_{j}{\lambda_j^{\zeta_{ij}}}.
\end{equation}
where $\lambda_j$ is the overall probability that a node is assigned to the $j$th class. Therefore the joint probability for $(\boldsymbol{x}_i, \boldsymbol{\zeta}_i)$ is given by the product 
\begin{equation}
\p(\boldsymbol{x}_i,\boldsymbol{\zeta}_i|\boldsymbol{\theta},\boldsymbol{\lambda})=\p(\boldsymbol{x}_i|\boldsymbol{\zeta}_i,\boldsymbol{\theta})\p(\boldsymbol{\zeta}_i|\boldsymbol{\lambda})=\prod_{j}{\left[\lambda_{j}\p(\boldsymbol{x}_i|\theta_j)\right]}^{\zeta_{ij}}.
\end{equation}
By marginalizing over all possible values of the indicator variables $\zeta_{ij}$, a \emph{mixture of Gaussians} model for the optical parameters is obtained
\begin{equation}
\p(\boldsymbol{x}_i|\boldsymbol{\theta},\boldsymbol{\lambda})=\int_{\boldsymbol{\zeta}_i}{\p(\boldsymbol{x}_i,\boldsymbol{\zeta}_i|\boldsymbol{\theta},\boldsymbol{\lambda}){\mathrm{d}\boldsymbol{\zeta}_i}}= \sum_{j}{\lambda_j\p(\boldsymbol{x}_i|\theta_j)}.
\label{equation_ch3_model_gaussmix}
\end{equation}
Finally, for independent nodes the prior of the image is given by
\begin{equation}
\p(\boldsymbol{x}|\boldsymbol{\theta},\boldsymbol{\lambda})=\prod_i \sum_{j}{\lambda_j\p(\boldsymbol{x}_i|\theta_j)}.
\label{eq_prior}
\end{equation}

\subsubsection{Reconstruction step}
The objective function takes the form of equation \eqref{eq_objfungen}, where at iteration $t$ of the reconstruction-classification algorithm the regularization is given by (equations \eqref{eq_reg} and \eqref{eq_prior})
\begin{equation}
\mathcal{R}^t(\boldsymbol{\mua},\boldsymbol{\musp}) = -\log{\p(\boldsymbol{x}|\boldsymbol{\theta}^t,\boldsymbol{\lambda}^t)} = \frac{\tau}{2} {\|\mathbf{L}_{\bar{x}}(\boldsymbol{x}-\bar{\boldsymbol{x}})\|}^2,
\label{eq_regul}
\end{equation}
where $\tau$ is a regularization parameter and
\begin{equation}
\bar{\boldsymbol{x}}_i=\left.\sum_{j}{ {\zeta_{ij}\cdot \boldsymbol{m}_{j}}}\right|_{\mathrm{MAP}(\boldsymbol{\zeta})}\!\!\!\!=\,\boldsymbol{m}_{j'}\ \ \ \ \ \ \in\mathbb{R}^2
\label{eq_xbar}
\end{equation}
is obtained by the fixing the labels to the \textit{maximum a posteriori} estimate, given the results of the previous iteration
\begin{equation}
\mathrm{MAP}(\boldsymbol{\zeta})=\arg\max_{\boldsymbol{\zeta}}\p(\boldsymbol{\zeta}|\boldsymbol{x}^{t-1}, \boldsymbol{\theta}^{t-1}, \boldsymbol{\lambda}^{t-1}),
\label{eq_maplabels}
\end{equation}
which is calculated in the classification step (see section \ref{section_class}). The weighting matrix $L_{\bar{x}}$ is the Cholesky decomposition of ${\Sigma_{\bar{x}}}^{-1}$, where $\Sigma_{\bar{{x}}}\in \mathbb{R}^{2N\times 2N}$ is a sparse matrix of which the $i$th $2\times 2$ block along the diagonal is $\Sigma_{j'}$ if the $i$th element belongs to the $j'$th class. 

In order to sphere the solution space, that is to render the space dimensionless, we performed a change of variables $\boldsymbol{\mua}\rightarrow\boldsymbol{\mua}/\boldsymbol{\mua}_{\,0}$ and $\boldsymbol{\musp}\rightarrow\boldsymbol{\mua}/\boldsymbol{\musp}_{\,0}$, where $(\boldsymbol{\mua}_{\,0},\boldsymbol{\musp}_{\,0})$ is the initial guess for the optical parameters (in this study, we initialized to the homogeneous background). Given the size of problem, we chose a gradient-based optimization method in order to reduce memory use and computational expense \cite{Saratoon2013}. The minimization was performed using the limited-memory Broyden-Fletcher-Goldfarb-Shanno (L-BFGS) method \cite{Nocedal1999}, with a storage memory of 6 iterations.

\subsubsection{Classification} \label{section_class}
The purpose of the classification step is to update the multinomial model, using the result of the previous reconstruction step. First, the expected values of the labels $\boldsymbol{\zeta}^{t+1}$ are computed for the current class parameters $(\boldsymbol{\theta}^t,\boldsymbol{\lambda}^t)$ and image $\boldsymbol{x}^t=(\boldsymbol{\mua}^t,\boldsymbol{\musp}^t)$ (E-step). Then the model parameters are updated by maximizing the posterior probability (M-step)
\begin{equation}
\p(\boldsymbol{\theta},\boldsymbol{\lambda}|\boldsymbol{x}^t) \propto \p(\boldsymbol{x}^t| \boldsymbol{\theta}, \boldsymbol{\lambda})\p(\boldsymbol{\theta},\boldsymbol{\lambda}).
\end{equation}
\\
\textbf{E-step}:\\
The \emph{responsibility} $r_{ij}^t$ is a measure of the probability that the $i$th node is assigned to the $j$th class. Using Bayes' theorem and the Gaussian mixture model \eqref{equation_ch3_model_gaussmix} we have
\begin{IEEEeqnarray}{rCl}
\p(\zeta_{ij}=1|\boldsymbol{x}_i^t,\boldsymbol{\theta}^t,\boldsymbol{\lambda}^t)&=&\frac{\p(\boldsymbol{x}_i|\zeta_{ij}=1,\boldsymbol{\theta}^t)\p(\zeta_{ij}=1)}{\p(\boldsymbol{x}_i|\boldsymbol{\theta},\boldsymbol{\lambda})}\nonumber\\
&=&\frac{\lambda_j^t\p(\boldsymbol{x}_i^t|\theta_j^t)}{\sum_j{\lambda_j^t \p(\boldsymbol{x}_i^t|\theta_j^t)}}=r_{nj}^t.
\label{eq_class_resp}
\end{IEEEeqnarray}
The expectation for the indicator values is 
\begin{IEEEeqnarray}{rCl}
E(\zeta_{ij}| \boldsymbol{x}_i^t, \boldsymbol{\theta}^t,\boldsymbol{\lambda}^t) &=& \int{\zeta_{ij}\p(\zeta_{ij}=1| \boldsymbol{x}_i^t,\boldsymbol{\theta}^t,\boldsymbol{\lambda}^t) \,\mathrm{d}\zeta_{ij}}\nonumber\\
&=&0\times\p(\zeta_{ij}=0|\boldsymbol{x}_i^t,\boldsymbol{\theta}^t,\boldsymbol{\lambda}^t)+1\times\p(\zeta_{ij}=1|\boldsymbol{x}_i^t,\boldsymbol{\theta}^t,\boldsymbol{\lambda}^t)\nonumber\\
&=&r_{ij}^t.
\end{IEEEeqnarray}
Therefore the MAP estimate for the labels is
\begin{equation}
	\zeta_{ij}^{t+1}= 
	\left\{
	\begin{array}{ll}
1 & \text{if }r_{ij}^t \text{ is maximum } \forall j,\\
0 &	\text{otherwise,}
	\end{array}
	\right.
\end{equation}
which can be used in equation \eqref{eq_maplabels}.
\\
\\
\textbf{M-step}:\\
The parameters $(\boldsymbol{\theta},\boldsymbol{\lambda})$ are chosen in order to maximize the log posterior	
\begin{equation}			
(\boldsymbol{\theta}^{t+1},\boldsymbol{\lambda}^{t+1})=\arg\max_{(\boldsymbol{\theta},\boldsymbol{\lambda})}{\log\p(\boldsymbol{x}^t | \boldsymbol{\theta}, \boldsymbol{\lambda})+\log\p(\boldsymbol{\theta},\boldsymbol{\lambda})}.
\end{equation}
Averaging over all possible values of $\boldsymbol{\zeta}$ gives
\begin{equation}
\log\p(\boldsymbol{x}^t | \boldsymbol{\theta}, \boldsymbol{\lambda})+\log\p(\boldsymbol{\theta},\boldsymbol{\lambda}) =\int_{\zeta}{\log\p(\boldsymbol{x}^t, \boldsymbol{\zeta} | \boldsymbol{\theta}, \boldsymbol{\lambda})\mathrm{d}\zeta}+\log\p(\boldsymbol{\theta},\boldsymbol{\lambda})
\end{equation}
Using \emph{Jensen's inequality} \cite{Prince2012} and ignoring terms which do not depend on $(\boldsymbol{\theta}, \boldsymbol{\lambda})$, we obtain a lower bound for the log-prior

\begin{align}
&\mathcal{B}(\boldsymbol{\theta},\boldsymbol{\lambda})=\sum_i\sum_j r_{ij}^t\log{(\lambda_j \p(\boldsymbol{\sigma}_{n}|\theta_j))}+\log\p(\boldsymbol{\lambda})+\log\p(\boldsymbol{\theta})\nonumber\\
&=\sum_i\sum_j r_{ij}^t\left[\log(\lambda_j)+\log(|\Sigma_{j}|)-\frac{1}{2} (\boldsymbol{x}_{i(n)}-\boldsymbol{m}_{j})'\Sigma_{j}^{-1}(\boldsymbol{x}_{i(n)}-\boldsymbol{m}_{j})\right]\nonumber\\
&+\sum_j{\left[(\alpha_{j}-1)\log(\lambda_j)-\frac{\nu_j+d+1}{2}\log{|\Sigma_{j}|}\right]}\nonumber\\
\end{align}

Maximizing $\mathcal{B}(\boldsymbol{\theta},\boldsymbol{\lambda})$ for $\sum_j\lambda_j=1$ and using non-informative priors, we obtain the update rules for the model parameters
\begin{equation}
\lambda_j^{t+1}=\frac{\sum_i r_{ij}^t}{N},
\label{eq_lamdba}
\end{equation}
\begin{equation}
\boldsymbol{m}_{j}^{t+1}=\frac{\sum_i r_{ij}^t\boldsymbol{x}_{i}}{\sum_i r_{ij}^t},
\label{eq_means}
\end{equation}
\begin{equation}
\Sigma_{j}^{t+1}=\frac{\sum_i r_{ij}^t (\boldsymbol{x}_{i}-\boldsymbol{m}_{j}) (\boldsymbol{x}_{i}-\boldsymbol{m}_{j})^T+\Gamma_j}{\sum_i r_{ij}^t+\nu_j+d+1}.
\label{eq_var}
\end{equation}

\begin{figure*}[t!]
\begin{minipage}{0.26\textwidth}
\centering
\subfloat[]{\includegraphics[width=.95\textwidth,trim=.in .in .in .in, clip=true]{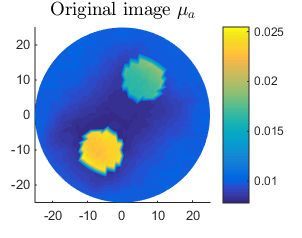} \label{fig_classinitorig}}

\vspace{.8cm}

\subfloat[] { \small
	\begin{tabular}{r|cc}
	 & $\mua$ & $\musp$\\
	\hline
	1&0.009&1.03\\
	2&0.024&0.15\\
	3&0.017&1.57 \vspace{0.5cm}\\
	\end{tabular}
	\label{fig_classinittable}}
	\normalsize
\end{minipage}
\begin{minipage}{0.74\textwidth}
\centering
\subfloat[]{\includegraphics[width=\textwidth,trim=.8in .2in .8in .in, clip=true]{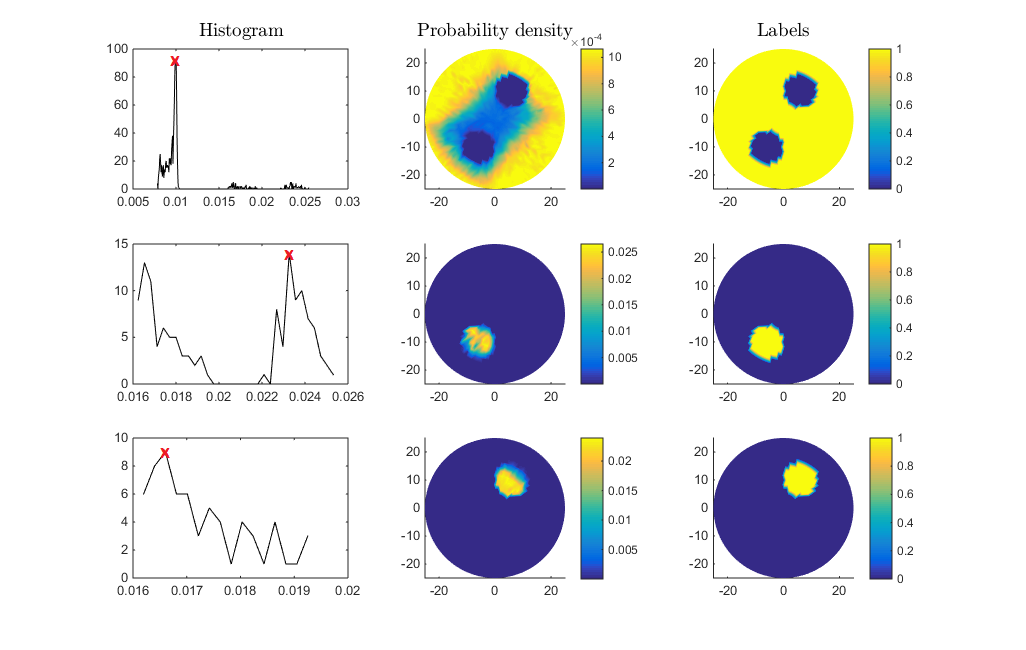}
\label{fig_classinitall}}
\end{minipage}
\caption{Class initialization example: (a) original image of $\mua$ to which we apply the segmentation; (b) result of taking average image values over the segmented areas (c) first column, histogram of occurrences of values of $\mua$ in the portion of the image requiring segmentation - value with highest number of occurrences is $\mua_h$ (indicated by a red cross); second column, probability density function with mean$(\mua_h,\musp_h)$ and covariance $\Sigma_h$; third column, labels identifying nodes with probability density higher than tolerance value $\mathrm{tol}_h$; each row corresponds to an iteration and a distinct class, so in this case $J=3$.}
\label{fig_classinit}
\end{figure*}

\subsection{Class means initialization} \label{section_classinit}
The number of classes $J$ and the class means $\boldsymbol{m}_j$ were initialized by automatically segmenting the result of the first reconstruction step and averaging over the segmented areas. To segment the image (for example, see figure \ref{fig_classinitorig}) we looked at a binned histogram of the image of $\boldsymbol{\mua}$ and chose the value $\mua_{\,h}$ for which the number of occurrences was highest (figure \ref{fig_classinitall}, column 1). We found the first node index $h$ for which the value $\mua_{\,h}$ occurs, and identified the corresponding scattering value $\musp_{\,h}$. Having chosen a covariance matrix $\Sigma_h$, we computed a map of the multivariate normal probability of the $(\boldsymbol{\mua},\boldsymbol{\musp})$ images, with mean $(\mua_{\,h},\musp_{\,h})$ (figure \ref{fig_classinitall}, column 2). Then we selected a tolerance level $\mathrm{tol}_h$ at which to truncate the probability map, and selected all nodes with probability higher than the tolerance as belonging to the same class as node $h$ (figure \ref{fig_classinitall}, column 3). We repeated this process on the remaining nodes until all nodes were classified. Thus the number of classes was set to the number of iterations, and the average of the optical parameters over each class was used to initialize the class means (figure \ref{fig_classinittable}).

\begin{figure}
\hrulefill
\begin{algorithmic}
\STATE Set $\mathrm{MaxIt}$, $\mathrm{tol}$
\STATE Initialize optical and class parameters $\boldsymbol{x}$, $\boldsymbol{\zeta}$, $\boldsymbol{\theta}$, $\boldsymbol{\lambda}$
\STATE Initialize iteration count $t=1$ and regularization term $\mathcal{R}^0 = 0$ (no regularization)

\REPEAT
\STATE \emph{Reconstruction:}

\STATE \ \ \ \ \ \ Update $\boldsymbol{x}$; minimize \eqref{eq_objfungen} using L-BFGS until $\mathcal{E}< \mathrm{tol}$
\IF{$t=1$}
\STATE \ \ Initialize class means $\boldsymbol{m}_j$ (section \ref{section_classinit})
\ENDIF
\STATE \emph{Classification:}
\STATE \ \ \ \ \ \ E-step; compute expected labels $\boldsymbol{\zeta}$ \eqref{eq_class_resp}
\STATE \ \ \ \ \ \ M-step; update class parameters $(\boldsymbol{\theta}$, $\boldsymbol{\lambda})$ \eqref{eq_lamdba}\eqref{eq_means}\eqref{eq_var}
\STATE Update regularization term $\mathcal{R}^t$ \eqref{eq_regul}
\STATE Re-initialize $\boldsymbol{x} = \bar{\boldsymbol{x}}$ \eqref{eq_xbar}
\STATE $t \gets t+1$
\UNTIL $t \geq \mathrm{MaxIt}$
\end{algorithmic}
\hrulefill
\caption{Reconstruction-classification algorithm outline}
\end{figure}

\subsection{Visualization of the results}
\begin{figure*}[t!]
\centering
\includegraphics[width=.87\textwidth,trim=1.9in .3in 1.45in .2in, clip=true]{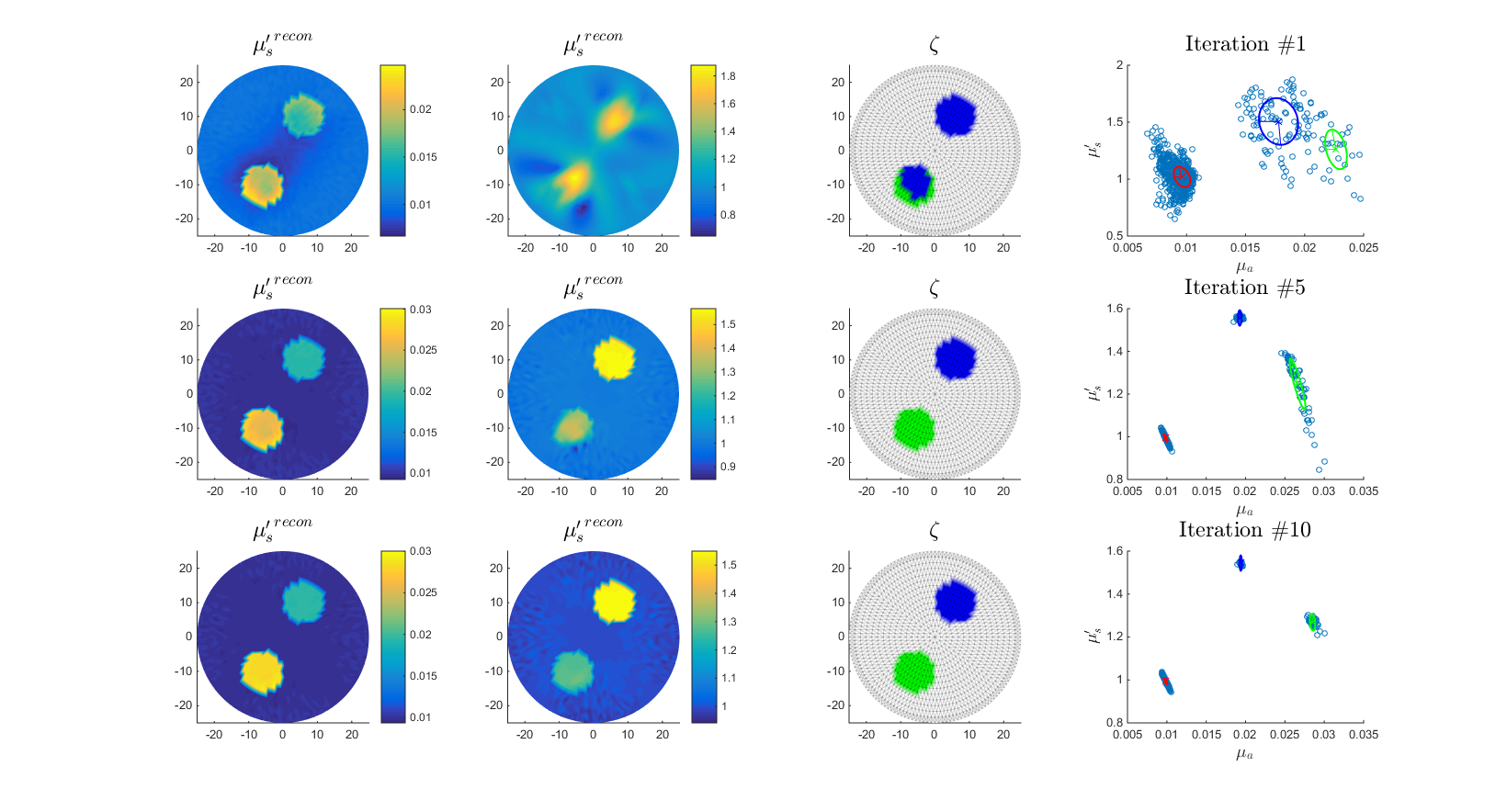}
\caption{2D reconstruction-classification results at iteration 1 (first row), 5 (second row) and 10 (third row). Reconstructed values of $\mua$ and $\musp$ (first and second column), labels recovered for perturbation classes (third and fourth columns), and scatter plot (fifth column).}
\label{fig_reconclass2D}
\end{figure*}

Results obtained using the reconstruction-classification method are displayed alongside scatter plots of the nodal values recovered in the 2D feature space $(\mua,\musp)$ (for example, see figure final column in \ref{fig_reconclass2D}). The positions of the class means $\boldsymbol{m}_j=(\bar{\mua}_j,\bar{\mus}'_j)$ are identified by a cross, and the class covariances $\Sigma_j$ are represented by ellipses. These are colour coded by class, and are indicative of the clustering of image nodal values around the class means.

\section{Results}

\begin{figure*}[t!]
\centering
\subfloat[]{
\includegraphics[width=0.16\textwidth,trim=.1in .in 1.in .2in, clip=true]{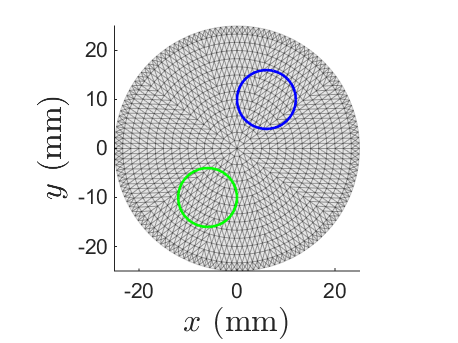} 
\label{fig_mesh}}
\hspace{2mm}
\subfloat[]{\includegraphics[width=0.80\textwidth,trim=1.5in 4.1in 1.5in 3.5in, clip=true]{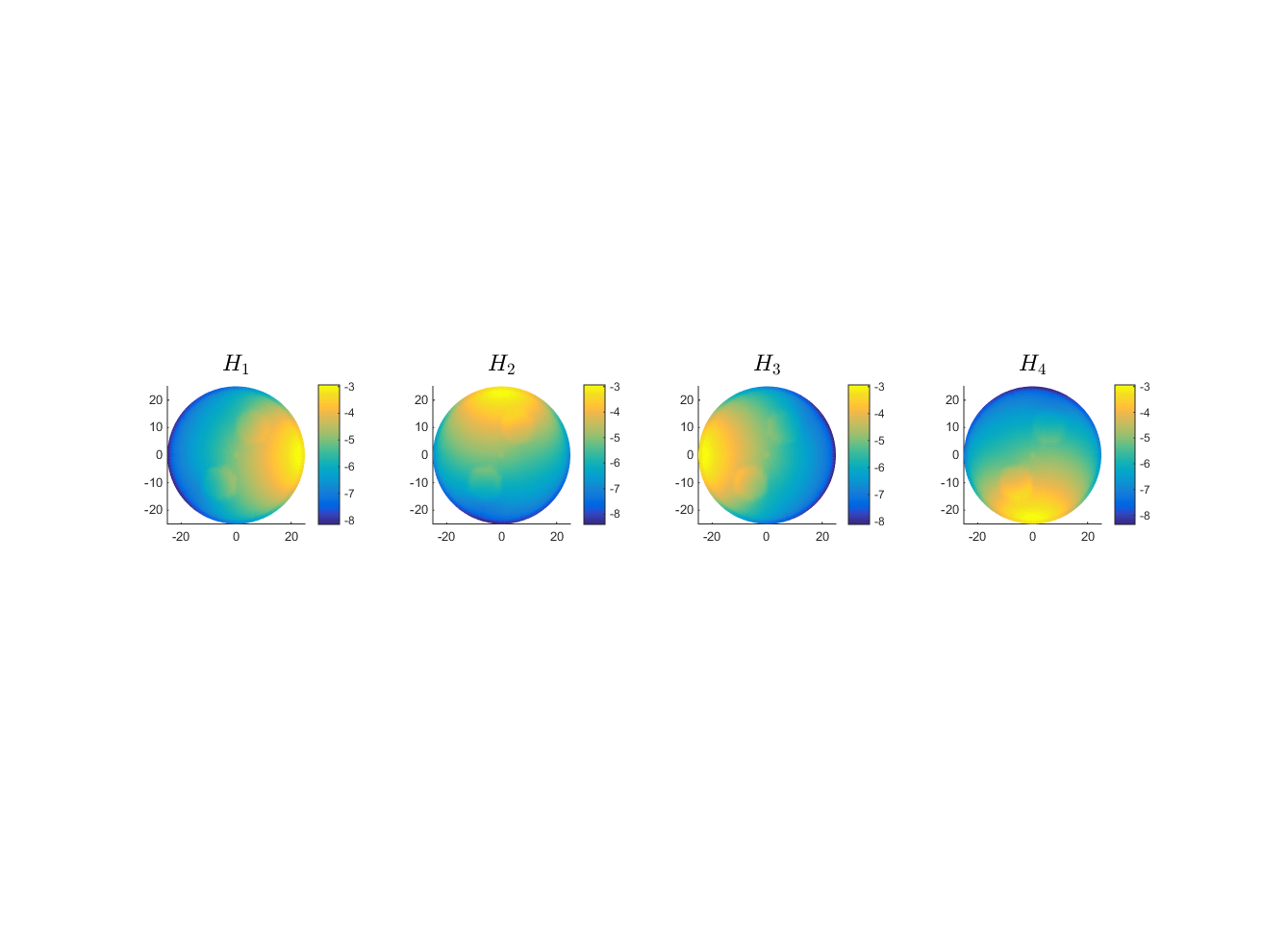}\label{fig_H2D}}
\caption{2D model: (a) circular mesh, (b) absorbed energy for each illumination pattern.}
\label{fig_model2D}
\end{figure*}

\begin{figure*}[t!]
\centering
\includegraphics[width=0.74\textwidth,trim=.8in .2in .8in .in, clip=true]{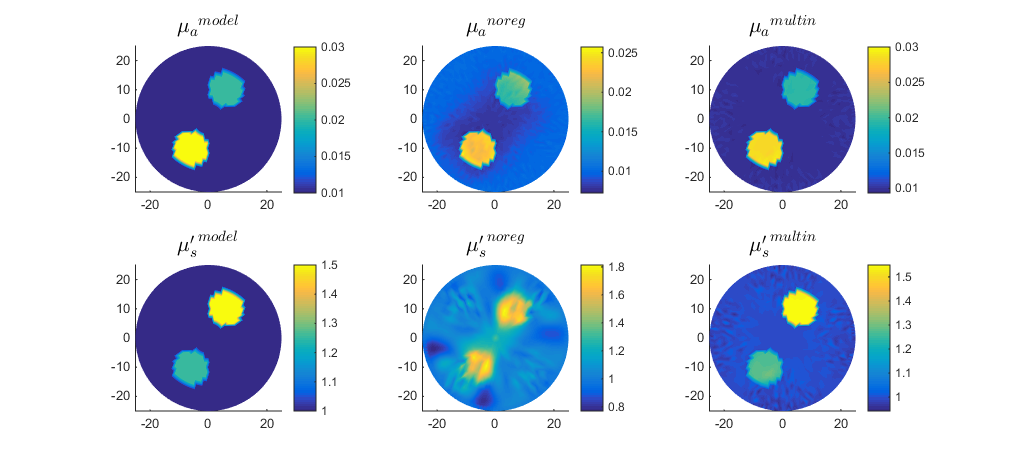}
\caption{2D model and reconstruction: first column, model of $\mus$ and $\musp$; second colunm: reconstructed values of $\mua$ and $\musp$ without multinomial prior; third column: reconstructed values of $\mua$ and $\musp$ with multinomial prior.}
\label{fig_modelrecon}
\end{figure*}

\subsection{2D validation and reconstruction}
We chose a numerical phantom defined on a 2D circular mesh with 1331 nodes and radius $25\,\mathrm{mm}$. Four illumination sources were placed on the boundary at angles $0$, $\pi/2$, $\pi$ and $3\pi/2\ \mathrm{rad}$. In all cases the illumination profile was a normalized Gaussian with radius (distance from the centre at which the profile drops to $1/e$) $6\,\mathrm{mm}$. The background optical parameters were set to $\mua = 0.01\,\mathrm{mm}^{-1}$ and $\musp = 1\,\mathrm{mm}^{-1}$. Two circular perturbations of radius $6\,\mathrm{mm}$ were added in positions $(6\,\mathrm{mm}, 10\,\mathrm{mm})$ and $(-6\,\mathrm{mm}, -10\,\mathrm{mm})$ (figure \ref{fig_mesh}). The values of the perturbations were $\mua = 0.02\,\mathrm{mm}^{-1}$, $\musp = 1.5\,\mathrm{mm}^{-1}$ and $\mua= 0.03\,\mathrm{mm}^{-1}$, $\musp = 1.25\,\mathrm{mm}^{-1}$, respectively. The absorbed energy field was simulated for each illumination and 1\% white Gaussian noise was added (figure \ref{fig_H2D}). The class covariances were initialized to 
\begin{equation}
\Sigma_j= \left(
\begin{array}{cc}
10^{-6} & 0\\
0 & 10^{-1}
\end{array}
\right)\ \ \ \forall j= 1,\ldots,3,
\label{eq_sigmaj}
\end{equation}
where the first variable was the absorption and the second was the reduced scattering. The parameters of the Jeffreys prior were set to $\Gamma_j= \Sigma_j\ \forall j$, $\nu(1)=1$ for the background class, and $\nu(2,3)=10$ for the perturbation classes. The number of classes and optical parameters were initialized using the class means initialization method (section \ref{section_classinit}) with $\mathrm{tol}_h=10^{-5}$ and $\Sigma_h=\Sigma_j$ \eqref{eq_sigmaj}, and the labels were initialized to 1 for the background class and zero for all other classes. The tolerance of the L-BFGS algorithm was set to $\mathrm{tol} = 10^{-11}$ and the total number of reconstruction-classification iterations was set to $\mathrm{MaxIt}=10$ (figure \ref{fig_reconclass2D}). The regularization parameter $\tau =10^{-10}$ was chosen by inspection. For comparison, images were reconstructed without introducing a prior (figure \ref{fig_modelrecon}); the images were reconstructed by minimizing \eqref{eq_objfunnoreg} using the L-BFGS method with $\mathrm{tol} = 10^{-12}$.

\begin{figure*}[t!]
\centering
\subfloat[]{
\includegraphics[width=0.25\textwidth,trim=.3in .in .1in .1in, clip=true]{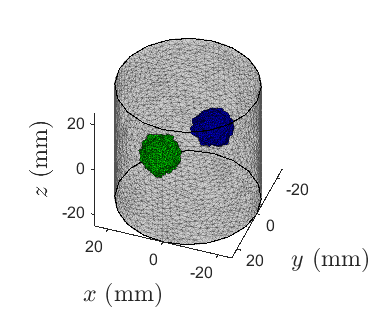} 
\label{fig_model3D}}
\subfloat[]{
\includegraphics[width=0.25\textwidth,trim=.3in .in .1in .1in, clip=true]{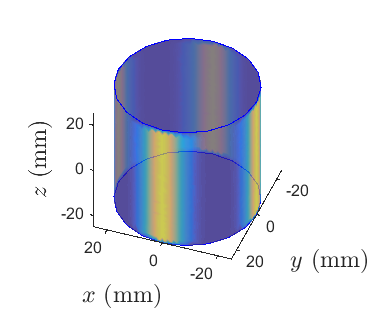} 
\label{fig_sources3D}}
\subfloat[]{
\includegraphics[width=0.45\textwidth,trim=.5in 1in .5in 1in, clip=true]{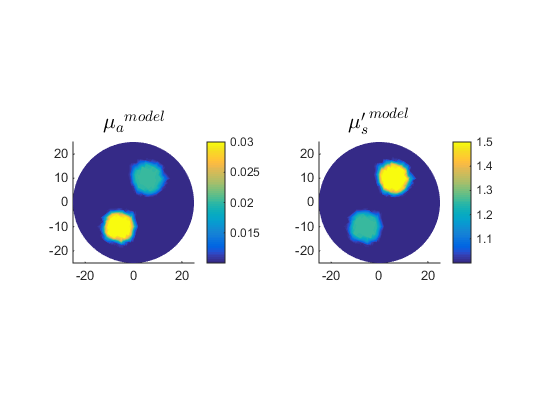} 
\label{fig_optmodel3D}}

\subfloat[]{
\includegraphics[width=0.9\textwidth,trim=1.6in 0in 1.6in 0.in, clip=true]{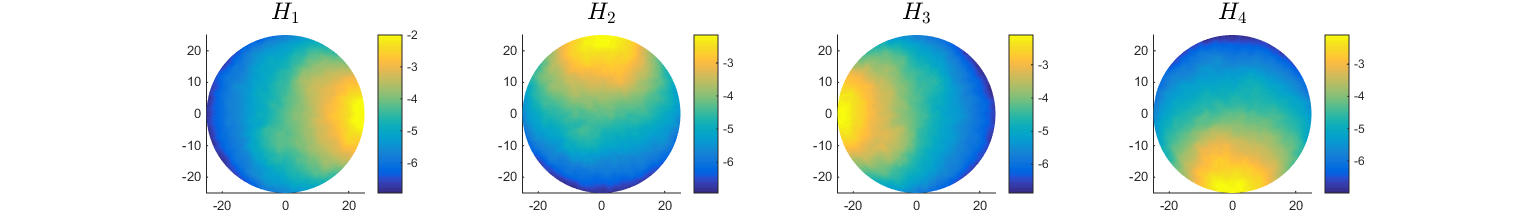}
\label{fig_H3D}}
\caption{3D model: (a) numerical phantom and perturbation locations, (b) all illumination sources, (c) cross section of optical parameters used to simulate the data for $z=0$, (d) cross section of absorbed energy for each illumination pattern.}
\label{fig_model3D_all}
\end{figure*}

\begin{figure*}[t!]
\centering
\includegraphics[width=.9\textwidth,trim=1.8in .2in 1.4in .2in, clip=true]{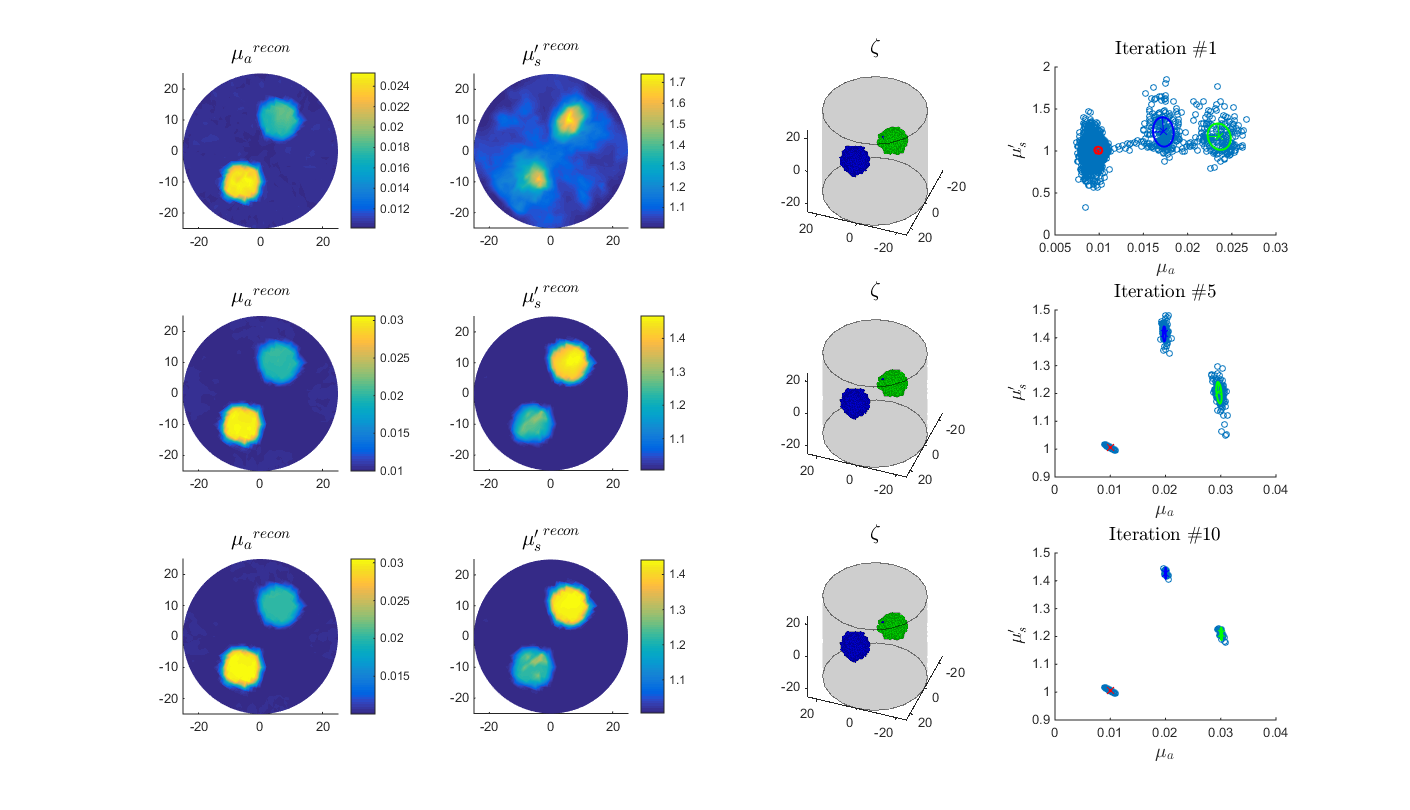}
\caption{3D reconstruction-classification results at iteration 1 (first row), 5 (second row) and 10 (third row). Reconstructed values of $\mua$ and $\musp$ (first and second column), labels recovered for perturbation classes (third column), and scatter plot (fourth column).}
\label{fig_reconclass3D}
\end{figure*}

\subsection{3D validation and reconstruction}
We chose a 3D phantom analogous to the 2D case, defined on a cylinder with 27084 nodes, radius $25\,\mathrm{mm}$ and height $25\,\mathrm{mm}$. Two spherical inclusions of radius $6\,\mathrm{mm}$ were placed in $(6\,\mathrm{mm}, 10\,\mathrm{mm}, 0\,\mathrm{mm})$ and $(-6\,\mathrm{mm}, -10\,\mathrm{mm}, 0\,\mathrm{mm})$ (figure \ref{fig_model3D}). Illuminations sources were Gaussian in the xy-plane constant in the z-axis, with radius $6\,\mathrm{mm}$ and length $25\,\mathrm{mm}$ (figures \ref{fig_sources3D}, \ref{fig_optmodel3D}). PAT images were simulated for 4 illuminations at the cardinal points, and 1\% noise was added to the absorbed energy (figure \ref{fig_H3D}). The optical, covariance and reconstruction parameters were set to the same values used in the 2D case. The class initialization parameters were set to $\mathrm{tol}_h=10^{-7}$ and $\Sigma_h=\Sigma_j$. Images were reconstructed by performing 10 iterations of the reconstruction-classification method (figure \ref{fig_reconclass3D}).

\section{Discussion}

\subsection{Summary of findings}
We applied the proposed reconstruction-classification algorithm to a 2D numerical phantom with 3 tissues, a background and 2 perturbations (figure \ref{fig_model2D}). The optical absorption was recovered reliably within a small number of iterations, and the scattering was recovered with sufficient accuracy after approximately 10 iterations (figure \ref{fig_reconclass2D}). We compared the optical model with images obtained by the reconstruction-classification method, and by a traditional reconstruction-only (no regularization) method (figure \ref{fig_modelrecon}). We found that the reconstruction-classification method delivered superior image quality, particularly with regards to the scattering parameter. We applied the reconstruction-classification algorithm to a much larger 3D problem (figure \ref{fig_model3D_all}) and observed similar results (figure \ref{fig_reconclass3D}) as in the 2D case.

\subsection{Choice of parameters}
The parametric optical model and classification algorithm introduce a number of parameters which require tuning by the user. In addition to the regularization parameter, the parameters of the Jeffreys prior $\Gamma$ and $\nu$ and the initial guess of the class variances $\Sigma_j$ must be set before performing the classification. However, their significance is fairly intuitive, and with experience of a certain type of problem the choice of parameters becomes natural. Visualizing the class covariance matrix $\Sigma_j$ as an ellipse, changing the value of $\Gamma$ varies its eccentricity, and changing $\nu$ varies the length of its axes. Further, given that in the first iteration the optical absorption is recovered with superior accuracy than the scattering, it is preferable to initialize the variance of the former to a smaller value than the latter, indicating greater confidence in the imaging solution.

\subsection{Initialization of the class means}
The purpose of the means initialization scheme is to increase automation of the method, so that minimum user intervention and no prior knowledge of the number of tissues or their optical properties is required. The algorithm simply performs a segmentation of the image, and then takes averages over the segmented areas to initialize the class properties (figure \ref{fig_classinit}). Alternative segmentation techniques could have been employed, however the advantage of the proposed approach is that it directly exploits the mixture of Gaussians model to identify the tissues. Our choice to investigate a node $h$ with $\mua$ belonging to the bin with maximum number of occurrences leads to the background tissue being identified first, followed by the perturbation tissues. The choice of the node index $h$ could have been randomized, so that tissues are identified in random order. This approach is equally valid, however we found that in cases where tissue values were close together (such as after a single reconstruction-classification iteration) it was preferable to identify the largest classes first because the mean was estimated with greater accuracy for the classes with a larger number of samples. Further, for a given image and tolerance level, our choice renders the result of the segmentation process unique and reproducible.

\subsection{Recovery of the scattering}
From the comparison with the reconstruction-only case with no regularization (figure \ref{fig_modelrecon}), it is evident that the introduction of the parametric prior enables better recovery of the scattering. The inconsistency between the quality of the recovered absorption and scattering parameters in the non-regularized case is due to the weaker dependence of the latter on the absorbed energy density with respect to the former. This results in the scattering gradient being approximately an order of magnitude smaller than the absorption gradient. Although the problem can be mitigated by sphering the solution space, variations in the data due to the scattering often fall below the noise floor. In the reconstruction-classification case, typically the absorption is recovered with good accuracy within a small number of iterations. Thus, the absorption takes values very close to the class means (resulting in small clusters), and the variance along the $\mua$ direction converges to a small value. Given that the regularization term is weighted by the inverse of the covariance matrix, the dependence of the absorption gradient on the data becomes weaker at each iteration, until its magnitude is comparable or smaller to that of the scattering. In the iterations that follow, the descent of the data term of the objective function is primarily due to updating the scattering, which converges to the correct values.  

\subsection{Computational demands}
Computational performance was found to be strongly dependent on the problem size. In the 2D case with 1331 nodes (figure \ref{fig_reconclass2D}), the total reconstruction time (10 outer reconstruction-classification iterations) using Matlab on a 16 core PC with 128GiB RAM was only 77 seconds. In the 3D case with 27084 nodes (figure \ref{fig_reconclass3D}), the total reconstruction time increased linearly with the number of nodes, and on the same workstation was approximately 3.7 hours. The increase in computation time was mostly due to much longer processing times for the L-BFGS algorithm in the reconstruction step.  

\subsection{Experimental application}
In experimental situations, prior information on tissue properties may be held, such as knowledge of the characteristic optical absorption and scattering spectra of chromophores of interest. These may be obtained from the literature \cite{Jacques2013}, or gained through tissue sample measurements. This information could be used in one of two ways. Firstly, a library of typical chromophores could be used to initialize the class parameters, instead of the proposed class means initialization method. The classification process could then perform the function of correcting for uncertainty, errors or local variations in the real optical properties with respect to the prior information. Alternatively, it could be used to label the chromophores found by the segmentation process, and identify these as certain tissues such as for example `oxygenated blood' or `fat', on the basis of the closeness of the recovered means to the characteristic properties.

\subsection{Additional priors}
In this study we assumed independence between nodal values, however the mixture of Gaussian model could be used in conjunction with a spatial prior. Knowledge of smoothness or sparsity properties of the solution could be employed to introduce a homogeneous spatial regularizer such as first-order Tikhonov \cite{Saratoon2013a} or Total Variation \cite{Bal2011,Tarvainen2012}. Knowledge of structural information, such as that provided by an alternative imaging method or anatomical library, could be exploited by introducing a spatially varying probability map for the optical properties.

\section{Conclusions}
In this paper, we proposed a novel method for performing image reconstruction in QPAT. We introduced a parametric class model for the optical parameters, and implemented a minimization-based reconstruction algorithm. We suggested an automated method by which to initialize the parameters of the class model, and proposed a classification algorithm by which to progressively update and improve those parameters after each reconstruction step. We demonstrated though 2D and 3D numerical examples that the reconstruction-classification method allows for the simultaneous recovery of optical absorption and scattering. In particular, we found that this approach delivered superior accuracy in the recovery of the scattering with respect to traditional gradient-based reconstruction.

\section*{Acknowledgements}
This work was funded by the EPSRC Doctoral Prize Fellowship EP/M506448/1.

\bibliography{qpatrc2d.bib}  
\bibliographystyle{elsarticle-num-nourl} 

\end{document}